\documentclass[twocolumn,aps,prc,superscriptaddress,showpacs]{revtex4}
\usepackage{amsmath,bm}
\usepackage{graphicx}
\usepackage{dcolumn}
\usepackage{multirow}
\newcolumntype{z}[1]{D{.}{.}{#1}}
\begin{document}
\draft

\title{Density  dependence of the ``symmetry energy" in the lattice gas model }

\author{Q. M. Su}
\affiliation{Shanghai Institute of Applied Physics, Chinese
Academy of Sciences, P. O. Box 800-204, 201800 Shanghai, People's
Republic of China} \affiliation{Graduate School of the Chinese
Academy of Sciences, 100080 Beijing, People's Republic of
ChinaChina} \affiliation{College of Electronic and Electrical
Engineering, Shanghai University of Engineering Science, 201620
Shanghai, People's Republic of China}
\author{Y. G. Ma}
 \thanks{ Corresponding author. Email: ygma@sinap.ac.cn }
\author{W. D. Tian}
\author{D. Q. Fang}
\author{X. Z. Cai}
\author{K. Wang}
\affiliation{Shanghai Institute of Applied Physics, Chinese
Academy of Sciences, P. O. Box 800-204, 201800 Shanghai, People's
Republic of China}
\date{\today}

\begin{abstract}
Isoscaling behavior of the statistical emission fragments from the
equilibrated sources with $Z$ = 30 and $N$ = 30, 33, 36 and 39,
resepectively, is investigated in the framework of isospin
dependent lattice gas model. The dependences of isoscaling
parameters $\alpha$ on source isospin asymmetry, temperature and
freeze-out density are studied and the ``symmetry energy" is
deduced from isoscaling parameters. Results show that ``symmetry
energy" $C_{sym}$ is insensitive to the change of temperature but
follows the power-law dependence on the freeze-out density $\rho$.
The later gives $C_{sym}$ = 30$(\rho/\rho_0)^{0.62}$ if the
suitable asymmetric nucleon-nucleon potential is taken. The effect
of strength of asymmetry of nucleon-nucleon interaction potential
on the density dependence of the ``symmetry energy" is dicussed.

\end{abstract}


\maketitle

\section{\label{sec:sec1}Introduction}

One of major goals of nuclear physics is to understand nature of
the nuclear matter, such as the nuclear equation of states (EOS)
and interaction of their basic builing blocks, namely neutrons and
protons. Especially, the propoerties of the asymmetric nuclear
matter is of great interesting in recent years since it has
important implications for astrophysics and nuclear physics
studies. The availability of beams with large neutron-to-proton
ratio, $N/Z$, provides the opportunity to explore the symmetry
energy term in nuclear equation of state in very asymmetric
nuclear systems. In such reactions, isospin degree of freedom has
a prominent role and can serve as a valuable probe of the symmetry
energy term of the nuclear equation of state. Recently the
isoscaling behavior extracted from the fragment isotopic
composition between two reactions with similar entrance channel
has been  proposed as a very powerful tool to determine the
symmetry energy term of EOS \cite{Tsang2001PRL}. This scaling law
relates ratios of isotope yields measured in two different nuclear
reactions, 1 and 2, $R_{21}(N,Z) = Y_2(N,Z)/Y_1(N,Z)$, and such
ratios are shown to obey an exponential dependence on the neutron
number $N$ or proton number $Z$ of the isotopes or isotones
characterized by three parameters $\alpha$, $\beta$ and $C$
\cite{Tsang2001PRL}:
\begin{equation}
R_{21}(N,Z) = \frac{Y_2(N,Z)}{Y_1(N,Z)} = C \exp(\alpha N + \beta
    Z),
\end{equation}
where $C$ is an overall normalization constant. So far, this
universal law expression for $R_{21}$ has been extensively
explored experimentally and theoretically, under diverse
approximations, to primary yields produced by disassembling
infinite equilibrated systems in microcanonical and grand
canonical ensembles as well in canonical ensembles. It has  been
observed in statistical models \cite{Tsang2001PRC,Bot,Tian2007},
lattice gas model \cite{Ma-lgm}, percolation model \cite{Dorso},
statistical abrasion-ablation model \cite{Fang,Zhong}, fission
model \cite{Ma-fission}, as well as the molecular dynamics type
models \cite{On2003PRC,Tian2006}. Furthermore, under these
approximations, the isoscaling parameters $\alpha$ and $\beta$
have been found to be related to the symmetry term of the nuclear
binding energy \cite{Tsang2001PRL,Bot}, to the level of isospin
equilibration \cite{Souliotis2004PRC}, and to the values of
transport coefficients \cite{Veselsky2004PRC}.

The paper presents results of the isoscaling analysis for the
light fragments from thermal sources which are produced by the
lattice gas model. The temperature and density dependences of of
the symmetry energy are extracted,  and the effect of interaction
potential between nucleon-nucleon on the symmetry energy is also
discussed.

The article is organized as follows. Section \ref{sec:sec2} makes
a simple introduction on the lattice gas model. In Section
\ref{sec:sec3}, isoscaling phenomenon and its dependence on
temperature and density are presented, while the temperature and
density dependence of the dedced ``symmetry energy" from the
isoscaling parameter has been determined.  The effect of
asymmertic nucleon-nucleon potential on the ``symmetry energy" is
also discussed. Finally a summary is given in Sec. IV.

\section{\label{sec:sec2}Model Overview}
The lattice gas model was developed to describe the liquid-gas
phase transition for atomic system by Lee and
 Yang \cite{Yang52}. The same model has already been applied to
nuclear physics for isospin asymmetrical systems
\cite{Jpan98,Ma99}. In the lattice gas  model, $A$ (= $N + Z$)
nucleons with an occupation number $s_i$ which is defined $s_i$ =
1 (-1) for a proton (neutron) or $s_i$ = 0 for a vacancy, are
placed on the $L$ sites of lattice. Nucleons in the nearest
neighboring sites interact with an energy $\epsilon_{s_i s_j}$.
The hamiltonian is written as
\begin{equation}
E = \sum_{i=1}^{A} \frac{P_i^2}{2m} - \sum_{i < j} \epsilon_{s_i
s_j}s_i s_j .
\end{equation}
In order to investigate the symmetrical term of nuclear potential
in this model, we can use sets of parameters: one is an attractive
potential constant  $\epsilon_{s_i s_j}$ between neutron and
proton but no interaction between like nucleon, i.e. proton and
proton or neutron and neutron, namely
\begin{eqnarray}
 \epsilon_{nn} \ &=&\ \epsilon_{pp} \ = \ 0. MeV \nonumber , \\
 \epsilon_{pn} \ &=&\ - 5.33 MeV.
\end{eqnarray}
This kind interaction results in an asymmetrical potential among
different kind of nucleon pairs, hence it is an isospin dependent
potential.
In previous studies, $\epsilon_{pn}$ is generally chosen to be
-5.33 MeV \cite{Jpan98,Ma99}. Later we will use different
$(\epsilon_{pn},\epsilon_{nn})$ parameters to compare the effect
of this asymmetrical potential on the isocaling parameters and
``symmetry energy".

In the LGM simulation, a three-dimension cubic lattice with $L$
sites is used. The freeze-out density of disassembling system is
assumed to be
 $\rho_f$ = $\frac{A}{L} \rho_0$, where $\rho_0$ is the normal
nuclear density. The disassembly of the system is to be calculated
at $\rho_f$, beyond which nucleons are too far apart to interact.
Nucleons are put into lattice by Monte Carlo Metropolis sampling.
Once the nucleons have been placed we also ascribe to each of them
a momentum by Monte Carlo samplings of Maxwell-Boltzmann
distribution. Once this is done the LGM immediately gives the
cluster distribution using the rule that two nucleons are part of
the same cluster if their relative kinetic energy is insufficient
to overcome the attractive bond, i.e. $P_r^2/2\mu - \epsilon_{s_i
s_j}s_i s_j < 0 $. This method is similar to the Coniglio-Klein's
prescription \cite{Coni80} in condensed matter physics. Then the
fragment isotopic composition can be acquired.

In the lattice gas model the ground state energy (binding energy)
can be defined as
\begin{equation}
 E_{bind}= \epsilon_{pn} N_{pn}
+ \epsilon_{nn} N_{nn}, \label{eq_Ebin}
\end{equation}
 where $N_{pn}$ or $N_{nn}$
is the number of the bonds between unlike nucleons or like
nucleons in the ground state which corresponds to a cold nucleus
at zero temperature and normal nuclear density in our present
classical model \cite{Ma_EPJA}. Practically, $N_{np}$ or $N_{nn}$
is determined by the geometry and is equal to the maximum bond
number of unlike nucleons or like nucleons of the systems.

\section{\label{sec:sec3}RESULTS AND DISCUSSIONS}

\subsection{Isoscaling behavior}

To make a systematic study of source parameters which might
influence the isoscaling behavior, several pairs of equilibrated
sources are considered at various initial temperature $T$ and
freeze-out density $\rho_f$ in our present work. The equilibrated
source pairs are chosen with different isospin asymmetry $N/Z$
with the fixed $Z$ = 30 but $A$ = 60, 63, 66 and 69, respectively,
which corresponds to $N/Z$ = 1, 1.1, 1.2 and 1.3, respectively. We
adopt the widely used convention to denote with the index
$^{\prime\prime}$2$^{\prime\prime}$ the more neutron-rich system
and with the index $^{\prime\prime}$1$^{\prime\prime}$ the more
neutron-poor system. In this situation the value of $\alpha$ is
always positive because more neutron-rich clusters will be
produced by the neutron-richer source and the value of $\beta$ is
always negative. The yield ratios $R_{21}(N,Z)$ are calculated and
the corresponding isoscaling behaviors are investigated over all
possible decayed fragments. The cubic lattices of $5^3$, $6^3$,
$7^3$, $8^3$ and $9^3$ are used for all systems, which gives the
freeze-out density $\rho_f$ = 0.48 $\rho_0$, 0.28 $\rho_0$, 0.18
$\rho_0$, 0.12 $\rho_0$ and 0.08 $\rho_0$, respectively, for $A$ =
60 system. There are slight different freeze-out densities for $A$
= 63, 66 and 69 systems. 100000 events were simulated for each
$T$-$\rho_f$ combination of each source which ensures good
statistics for results. In the present study, we will focus on the
light fragment isotopic distribution to explore the isoscaling
behavior and symmetry energy. The advantage of the LGM is that the
source density and temperature parameters are well defined.

\begin{figure}
\includegraphics[width=0.5\textwidth]{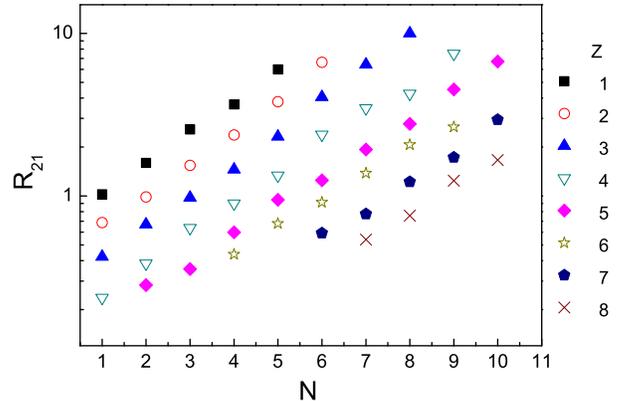}
\caption{\label{fig:fig1} (Color online) The ratio of the isotopes
for $Z < 9$ from two different sources for  $A$ = 66 and 63 with
$Z$ = 30
 at $T$ = 3  MeV and $\rho_f = 0.48 \rho_0$ as a function of the
 neutron number $N$. }
\end{figure}

Fig. \ref{fig:fig1} shows an example of the isoscaling behavior.
It plots the $R_{21}$ as a function of fragment neutron number $N$
from which the slope parameter (isoscaling parameter) can be
extracted. Average isoscaling parameter $\alpha$ which is
calculated over the range $Z \le 9$ is used to discuss the
dependence of $\alpha$ on the properties of emission source, such
as temperature, source size and source asymmetry of the isospin.
In the following calculations, isoscaling parameters $\alpha$
refer to the calculated average values.

\subsection{Temperature and density dependence of symmetry energy}

\begin{figure}
\includegraphics[width=0.5\textwidth]{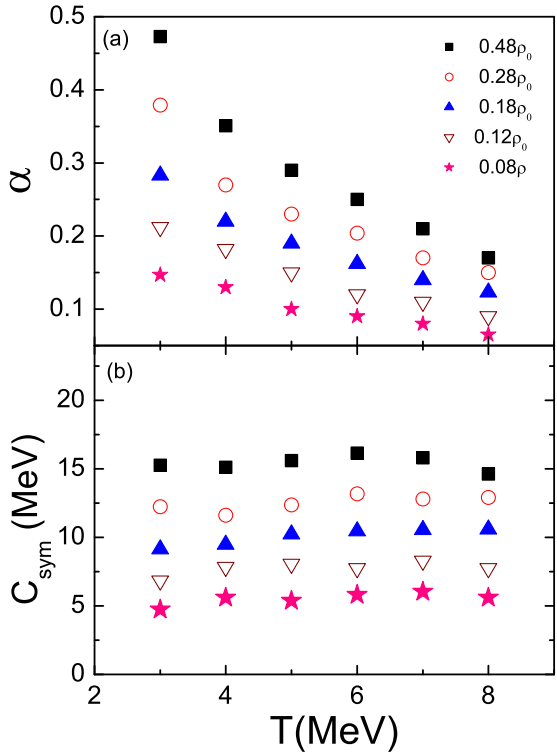}
\caption{\label{fig:fig2} (Color online) (a) Temperature and
density dependences of isoscaling parameters $\alpha$ extracted
from source pair with $A$ = 63 and 60. (b) Same as (a) but for the
deduced ``symmetry energy".}
\end{figure}

To explore the origin of isoscaling behavior and the dependence of
isoscaling parameters $\alpha$  on the isospin composition, we
performed calculations on source systems with different densities
and source asymmetry ($N/Z$) values. The temperature and density
dependences of $\alpha$  from the source pair ($N$ = 33, $Z$ = 30)
and ($N$ = 30, $Z$ = 30) are shown in Fig. \ref{fig:fig2}(a). Here
the scattering points are the calculation results for the LGM with
asymmetrical nucleon-nucleon potential, i.e. Eq.(3). Note that the
similar behavior for other source pairs was obtained but not shown
here. From the figure, $\alpha$ decreases as the temperature
arises, which indicates that the isospin dependence of the
fragment yields becomes weaker with the increasing of temperature.
But an increasing trend of $\alpha$ with density is also clearly
seen from the figure, which shows that lower density system has
lower chemical potential that makes smaller yield probable.

\begin{figure}
\includegraphics[width=0.45\textwidth]{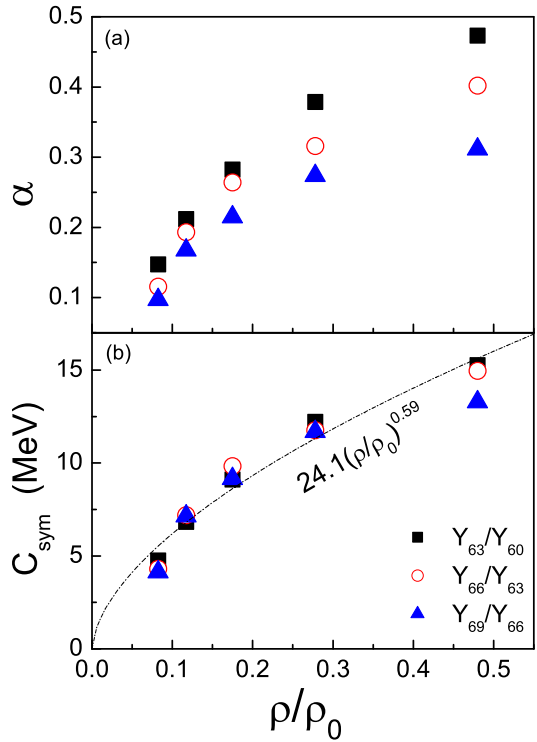}
\caption{\label{fig:fig3} (Color online) (a) $\alpha$ as a
function of freeze-out density for various source pairs with  $Z =
30$ which is illustrated in the figure at temperature $T$ = 3 MeV.
(b) The deduced ``symmetry energy" as a function of freeze-out
density. The line represents the fits with $C_{cym} = 24.1
(\rho/\rho_0)^{0.59}$. Symbols in figure correspond to
$Y_{A=63}/Y_{A=60}$ (solid square), $Y_{A=66}/Y_{A=63}$ (open
circles) and  $Y_{A=69}/Y_{A_s=66}$ (solid triangles).}
\end{figure}

It has been shown in the framework of the grand-canonical limit,
$\alpha$ has the form,
\begin{equation}
\alpha = \frac{4C_{\text{sym}}}{T}\Delta[(\frac{Z}{A})^2]=
\frac{4C_{\text{sym}}}{T}[(\frac{Z}{A})_1^2-(\frac{Z}{A})_2^2],
\end{equation}
where $C_{\text{sym}}$ is the deduced ``symmetry energy"
coefficient (MeV), $(\frac{Z}{A})_{\text{i=1,2}}^2$  means the
square of $Z$ over $A$ for system 1 and 2. $T$ is the temperature
of the system in MeV. Above equation has been proved to be a good
approximation by many calculations and experimental data, and are
generally adopted to constrain the symmetry energy coefficient
$C_{sym}$ in experiments. This behavior is attributed to the
difference of isospin asymmetry between two reaction systems in
similar nuclear temperature. Since the symmetry energy has an
important role on nuclear structure of neutron-rich or
neutron-deficient rare isotopes, studies on the isoscaling
behavior can be used to probe the isospin dependent nuclear
equation of state. Fig. \ref{fig:fig2}(b) shows that the deduced
``symmetry energy" $C_{sym}$ almost keep flat regardless the
change of temperature. In other words, the ``symmety energy" shows
little dependence on temparture. This is consistent with the
analysis of the symmetry energy versus the temperature within a
degenerate Fermi gas model \cite{Li06}. Furthermore, the deduced
``symmetry energy" displays an increasing trend with the
freeze-out density, which will be discussed later.

Fig. \ref{fig:fig3}(a) shows that the $\alpha$ increases with the
density for the three source pairs at $T$ = 3 MeV. If the Eq.~(4)
is satisfied in our simulation, $\alpha$ has a linear dependence
on $\Delta[(\frac{Z}{A})^2]$. Since this parameter is dependent on
the source systems, we divide $\alpha$ by
$\Delta[(\frac{Z}{A})^2]$ to remove the system isospin and size
dependences. Furthermore, since we know value of temperature for
each simulation, we can use the Eq.(4) to extract ``symmetric
energy" parameter. Fig.~\ref{fig:fig3}(b) shows the ``symmetric
energy" parameter $C_{sym} = \alpha T/[4 \Delta(\frac{Z}{A})^2]$
extracted from the ratios of different reaction system pairs,
namely $Y_{A=69}/Y_{A_s=66}$, $Y_{A=66}/Y_{A=63}$, and
$Y_{A=63}/Y_{A=60}$, as a function of density. Obviously, the same
dependence is displayed even though for the different source
combinations. Considering that the generally used
density-dependent symmetry energy coefficient has been taken,
namely
\begin{equation}
C_{cym} = C_0 (\rho/\rho_0)^\gamma,
\end{equation}
we can use the above expression to fit $C_{sym}(\rho)$. The line
in the Fig.~\ref{fig:fig3}(b) shows the fit. It gives $C_0 \sim
24$ MeV and $\gamma \sim 0.6$. It was worthy to mention that
$\gamma$ value is in a reasonable range, namely 0.55-0.69, which
was proposed by recent data and models \cite{Shetty}, but the
$C_0$ looks softer than the recent suggested value \cite{Shetty},
namely 31-33 MeV, even though it is consistent with the expanding
emitting source  model calculatiopn by Tsang et al.
\cite{Tsang2001PRL}. However, it must be mentioned that the
binding energy per nucleon $E_{bind}/A$  from the
Eq.~\ref{eq_Ebin} for the studied system in our LGM with
$(\epsilon_{pn} = -5.33,\epsilon_{nn}= 0)$ is around -11.7 MeV
which is greater than -16 MeV, the energy per nucleon in a uniform
nuclear matter at $T$ = 0 at the normal density $\rho_0$. Later we
will find the potential parameter set $(\epsilon_{pn} =
-7,\epsilon_{nn}= 0)$ will give a more reasonable binding energy
as well as $C_0$.

\begin{figure}
\includegraphics[width=0.5\textwidth]{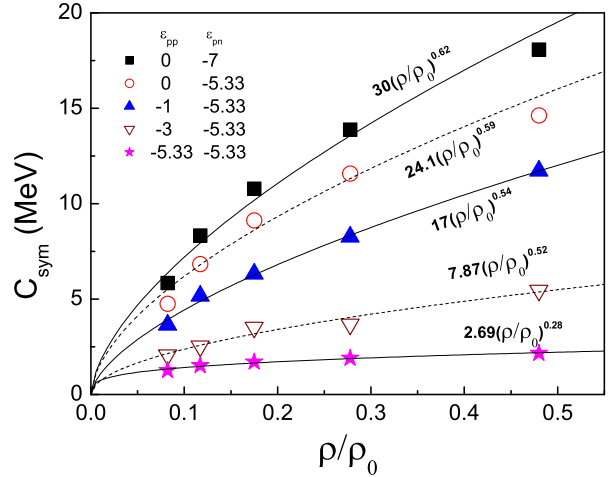}
\caption{\label{fig:fig5} (Color online) Deduced ``symmetry
energy" for ($N$ = 33, $Z$ = 30) and ($N$ = 30, $Z$ = 30) source
pairs at $T$ = 3 MeV as a function of density with different
($\epsilon_{pn}$, $\epsilon_{pp}$) parameters. The detailed values
for $\epsilon_{nn}$ and $\epsilon_{pn}$ are indicated in the
figure. Lines represent the fits with $C_{sym} = C_0
(\rho/\rho_0)^\gamma$ and the corresponding detail is illustrated
in the figure.}
\end{figure}

\subsection{Symmetry energy and nucleon-nucleon symmetric potential}

In order to discuss the effect of the isospin dependent
interaction potential on the deduced ``symmetry energy", we plot
Fig. \ref{fig:fig5} for $T$ = 3 MeV, by changing the symmetric
potential, i.e. on the one hand we change $\epsilon_{pp}$ from 0,
-1, -3 to -5.33 MeV but keep $\epsilon_{pn}$ = -5.33 MeV
constantly; on the other hand we change ($\epsilon_{pn}$ = -7,
$\epsilon_{pp}$ = 0) to comparison with the normal case, namely
$\epsilon_{pn}$ = -5.33 and $\epsilon_{pp}$ = 0. In fact, these
five ($\epsilon_{pn}$, $\epsilon_{pp}$) configurations, namely
from (-7,0), (-5.33,0), (-5.33,-1), (-5.33,-3) till (-5.33,-5.33),
counterparts gradually decreased asymmetric degree which can
characterized by the value of $\epsilon_{pn}$-$\epsilon_{pp}$.
When the symmetric potential becomes not sensitive slowly to
nucleus type, that is close to homogeneity. It is observed that
the isoscaling property is preserved for all situations, but the
values of the deduced ``symmetry energy" parameters shows
different sensitivity to the freeze-out density in different
condition of ($\epsilon_{pn}$, $\epsilon_{pp}$). As derived in
other models and theoretical frame \cite{Tsang2001PRL,
Tsang2001PRC}, isotope yield ratio is dominantly determined by the
symmetry term in the binding energy for two equilibrium sources
with comparable mass and temperature but different isospin degree
$N/Z$. It is clear that the values of the deduced "symmerty
energy"  in the isoscaling relation are related to the
¡°symmetry¡± potential between nucleon and nucleon: when
nucleon-nucleon has smaller symmetry potential (i.e. smaller
difference between $\epsilon_{pn}$ and $\epsilon_{pp}$) of the
system, the value of $C_{sym}$ decreases. To quantitatively see
the density dependence of ``symmetry energy", we fit the points
with the Eq.(6). The parameters of $C_0$ and $\gamma$ are shown in
the figure. Generally, both $C_0$ and $\gamma$ increase with the
asymmetry of the nucleon-nucleon potential. In other words, the
``symmetry energy" parameter $C_0$ tends to increase and
asymmetric EOS becomes harder with the asymmetry of the
nucleon-nucleon potential. It is of very interesting to see $C_0
\sim$ 30 MeV and $\gamma \sim$ 0.62 for the potential parameter
set $(\epsilon_{pn} = 7,\epsilon_{nn}= 0)$. The above parameters
are exactly consistent with the recently suggested form of density
dependence of the symmetry energy \cite{Shetty}. This is not
accidental since the LGM with the potential parameter set
$(\epsilon_{pn} = 7,\epsilon_{nn}= 0)$ gives a reasonable binding
energy ~-15.4 MeV/A for the studied systems, which is close to
the energy in a uniform nuclear matter at $T$ = 0 at the normal
density $\rho_0$.

\section{Conclusion}

In summary, the isotopic effect of  the light fragments from the
equilibrated thermal sources for the source pairs $N$ = 30, 33, 36
and 39 with constant $Z$ = 30 has been successfully investigated
by the lattice gas model model. The model constrains the source
temperature  and fragmenting density which is below saturate
nuclear density $\rho_0$, and the fragmentation starts from an
equilibrated system which is separated from any dynamical process.
In our work we change input parameters of temperature and
freeze-out density to survey the evolution of isospin degree of
freedom on the isoscaling behavior and the symmetry energy.
Isoscaling phenomenon is observed for the emitted light fragments
and their dependences on source isospin asymmetry, temperature and
density have been systematically investigated. Isoscaling
parameters $\alpha$ decrease with the increasing of temperature as
well as  the decreasing of density. The deduced ``symmetry energy"
$C_{sym}$ was extracted  from the isoscaling parameters, which
shows little dependence on temperature but follows the power-law
dependence on the freeze-out density, namely $C_{sym} = C_0
(\rho/\rho_0)^\gamma$. It is shown that $C_0 \sim$ 30 MeV and
$\gamma \sim $ 0.62 in case that the asymmetry nucleon-nucleon
potenatial ($\epsilon_{pn}$ = -7, $\epsilon_{pp}$ = 0) is used,
this result is consistent with the recently suggested values from
data and models. Finally the effect of the asymmetry
nucleon-nucleon potenatial on the deduced ``symmetry energy" are
studied by adjusting ($\epsilon_{pn}$, $\epsilon_{pp}$)
parameters, it shows that $C_0$ becomes larger and the density
dependence of symmetry energy becomes stiffer with the increasing
of asymmetric part of the nucleon-nucleon potential. Therefore it
is a powerful tool to use the isoscaling method to extract the
asymmetry EOS from the confrontation of data with the models.

This work was supported in part by the Shanghai Development
Foundation for Science and Technology under Grant Nos. 06JC14082
and 06QA14062, the National Natural Science Foundation of China
(NNSFC) under Grant Nos. 10535010 and 10775167, and the National
Basic Research Program of China (973 Program) under Contract No.
2007CB815004.

{}
\end{document}